# Melting dislocation transition in composite lithium greases with copper(II) carboxylate mesogenic additives

L. V. Elnikova[1], A. T. Ponomarenko[2], V. G. Shevchenko[2]

[1]NRC "Kurchatov Institute" – Alikhanov Institute for Theoretical and Experimental Physics, 117218 Moscow, Russian Federation
[2]Enikolopov Institute of Synthetic Polymeric Materials, RAS, 117393 Moscow, Russian Federation

We propose the theoretical description for the temperature and concentration phase transformations in composite lithium grease doped with the mesogenic additives of the copper(II) carboxylates family, copper(II) valerate and isovalerate, in the concentration range 1, 5, 10, 20 wt. %. The changes in character of specific electric conductivity of these composites manifested in experiments on dielectric spectroscopy in the measuring electric field of the frequencies 100 Hz – 1 MHz and polarization microscopy, are reliable caused by reversible phase transitions from the discotic phase to isotropic one or at heating from room temperature to 391 K. We study correlations between defect structure and electrical properties of these composites using the model of the Berezinsky-Kosterlitz-Thouless (BKT) transition. This transformation from the discotic to isotropic phase is associated with the evolution of topological defects dislocations induced by the mesogenic additives. The screw, transverse edge and longitudinal dislocations are realized in this case. We apply the numerical Monte Carlo technique to define their critical properties in the BKT transition.

**1. Introduction**

Lubricant compositions with additives are widely used in mechanical engineering due to their improved tribological performance compared to non-additive lubricants. The efficiency of such composites is exhibited in a decrease of the friction coefficient and wear, in an increase of machine parts life depending on mesogenic nature of additives [1–3].

Recently, in dielectric spectroscopy experiments, we confirmed that the electrical properties of synthetic greases of the Litol-24 family containing mesogenic additives, copper (II) carboxylates, valerate and isovalerate, change as a function of temperature and concentration (we studied them in the range 1, 5, 10, 20 wt%) [4].

To predict the behavior of a lubricant under operating conditions, we need to describe their phase states. For instance, if the mesogenic molecules forming a discotic phase oriented by their plane parallel to the friction surface, it is the best way to reduce the friction coefficient [2, 3]. The tribological testing of these lubricants in various loading conditions was carried out by Akopova and Terentyev [2,3]. The authors [2,3] concluded on the occurrence of mesomorphic transformations with a change in physical conditions and composition. When heated by 391 K, Cu(II) isovalerates form discotic phases [5], and the whole Litol-24 based composition, as expected, have to undergo the series of phase transformations. Also, in frames of the task of identification of the crystal symmetry, both thermotropic and lyotropic phases (i.e. concentration) transitions have to be considered, then knowing the thermodynamic parameters of the grease we can define the optimal amount of additives for the operation regimes.

From 70$^{\text{th}}$ years of the last century, there are well developed theoretical approaches in frames of the Landau-de Gennes theory and the theory of elasticity, which describe formation of discotic, columnar phases, such works were carried out by Bouligand [6], Kleman and Oswald [7–10], Kats and Monastyrsky [11,12], also we know many others similar theoretical approaches and numerical simulations [13–15]. The composite greases with columnar phases have also to possess topological defects that cause their mesomorphism. For the first time, Trebin proved [16] that the phase transitions in such systems are caused by dislocations.

On the other hand, the optimal modelling for formation of dislocations is just the two-dimensional theory of Berezinsky, Kosterlitz and Thouless (BKT) [17–20], as in its frames, it is possible to connect the electric values of the material and defect structure, i.e. macroscopic elasticity. In our paper, we use the BKT theory, as we provided our dielectric spectroscopy measurements for the unique lithium grease compositions, so that we can check the BKT approach independently and use the predictions of our modelling.

The second motivation of our theoretical analysis is to estimate experimental Frank moduli of our lithium grease compositions and to provide connection between such coefficients of the different sorts of elastic theories, *videlicet*, between the third order elasticity theory [10] and the first order Young's BKT approach [20] for dislocations in discotic phase.

## 2. Experimental

2.1. Materials

The basic multicomponent grease composition "Litol-24" (manufactured at the firm "OILRIGHT" [21]) is composed of the grease oil (30-40 %, molar mass 250…1000 g/mole), residual de-embedding oil (50-60 %, molar mass 390 000 g/mole), 12-lithium hydroxystearate (10-15%, molar mass 306.41 g/mole), diphenylamine (by 0.5 %, molar mass 169 g/mole), the additives are copper(II) isovalerate at concentrations 1, 5, 10, 20 wt. %.

Molar mass of $Cu(C_4H_9COO)_2$ is 435,546 g/mole, and the structure formula is shown in Fig.1. These additives are discotic mesogens [2, 3].

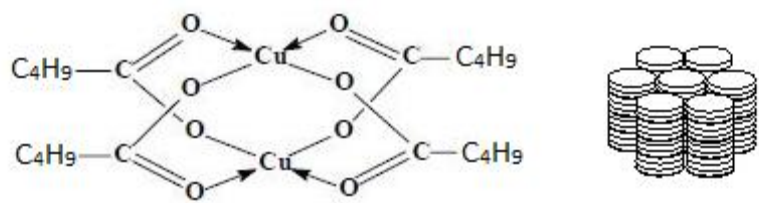

Fig. 1. Structure formula of Cu(II) carboxylates forming the columnar phases in composite lithium grease.

2.2. Polarization light microscopy

We recorded the series of microphotograms of the Litol-24's compositions using the method of polarization light microscopy at the microscope Axioscop 40 A Pol, Carl Zeiss, Germany at the temperature range from room temperature to 513 K at magnifications of 200x.

The microphotograms were analyzed with the ImageJ software (http://imagej.nih.gov/ij/docs/guide). The examples of forming of the hexatic phase and aggregates in the Litol-24's-Cu(II) isovarelate compositions are shown in Fig. 2. In Fig. 3, we show the distribution of mesogen particle sizes in cooling regime.

At heating and cooling, the Litol-24's compositions with different concentrations of Cu(II) valerate and isovalerate form multiphase regions, which reveal temperature hysteresis. And Litol-24-Cu(II) valerate compositions do not form a hexatic phase, but the crystalline phases only (Fig. 4).

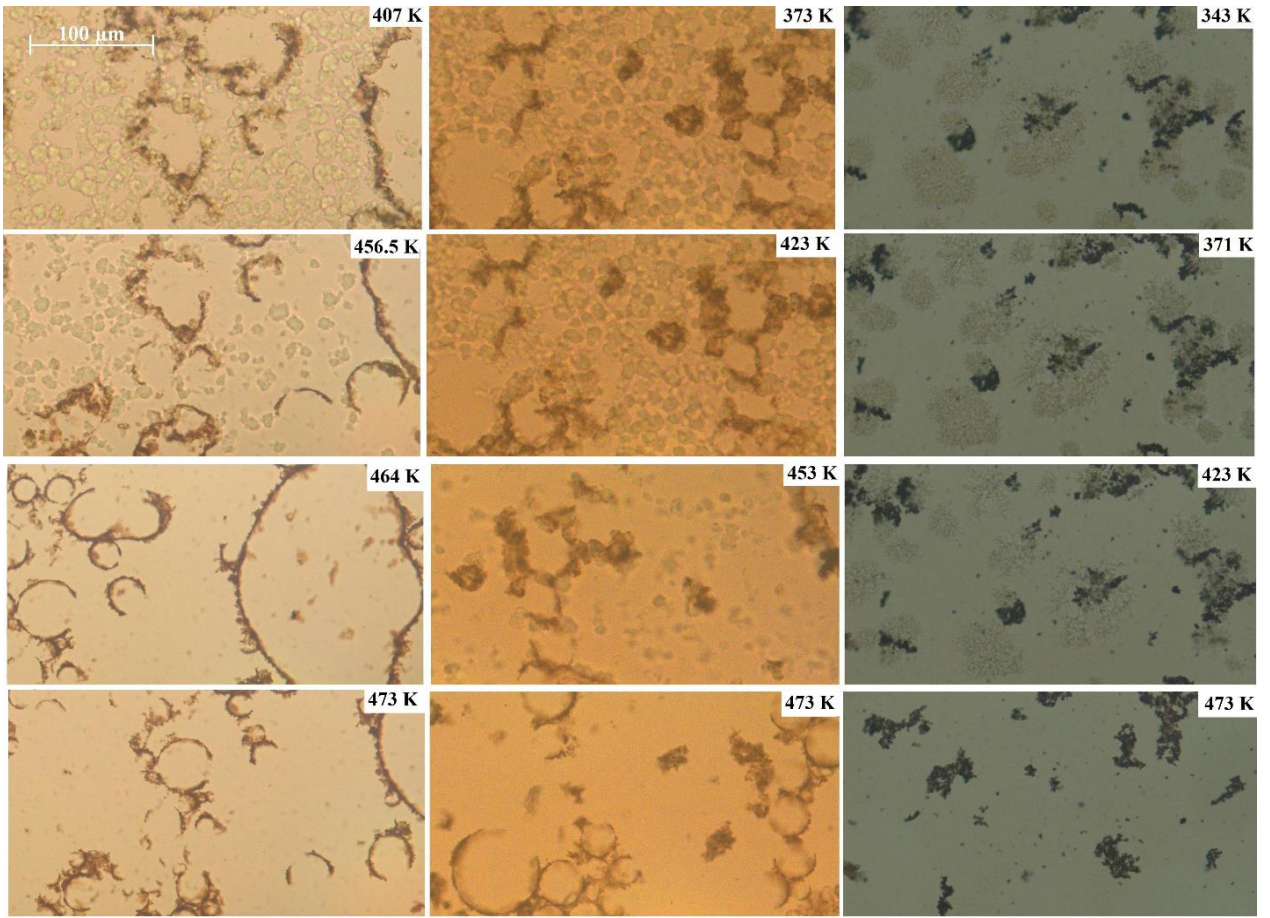

Fig. 2. Microphotographs of the system Litol-24-Cu(II) isovalerate 5, 10, 20 wt. % (left, middle and right columns respectively) at different temperatures in cooling regime.

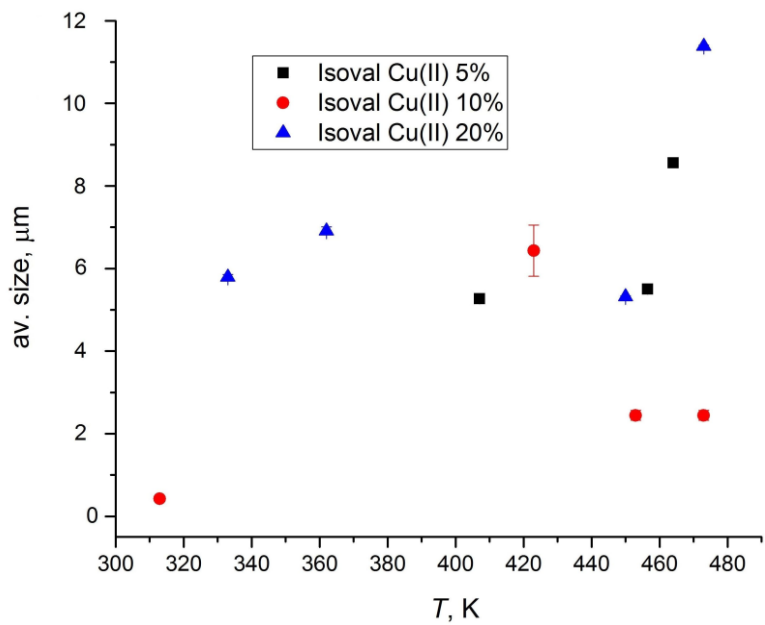

Fig. 3. Average sizes of Cu(II) isovalerate particles (at concentrations 5, 10, 20 wt.%) and aggregates in Litol-24 in cooling regime.

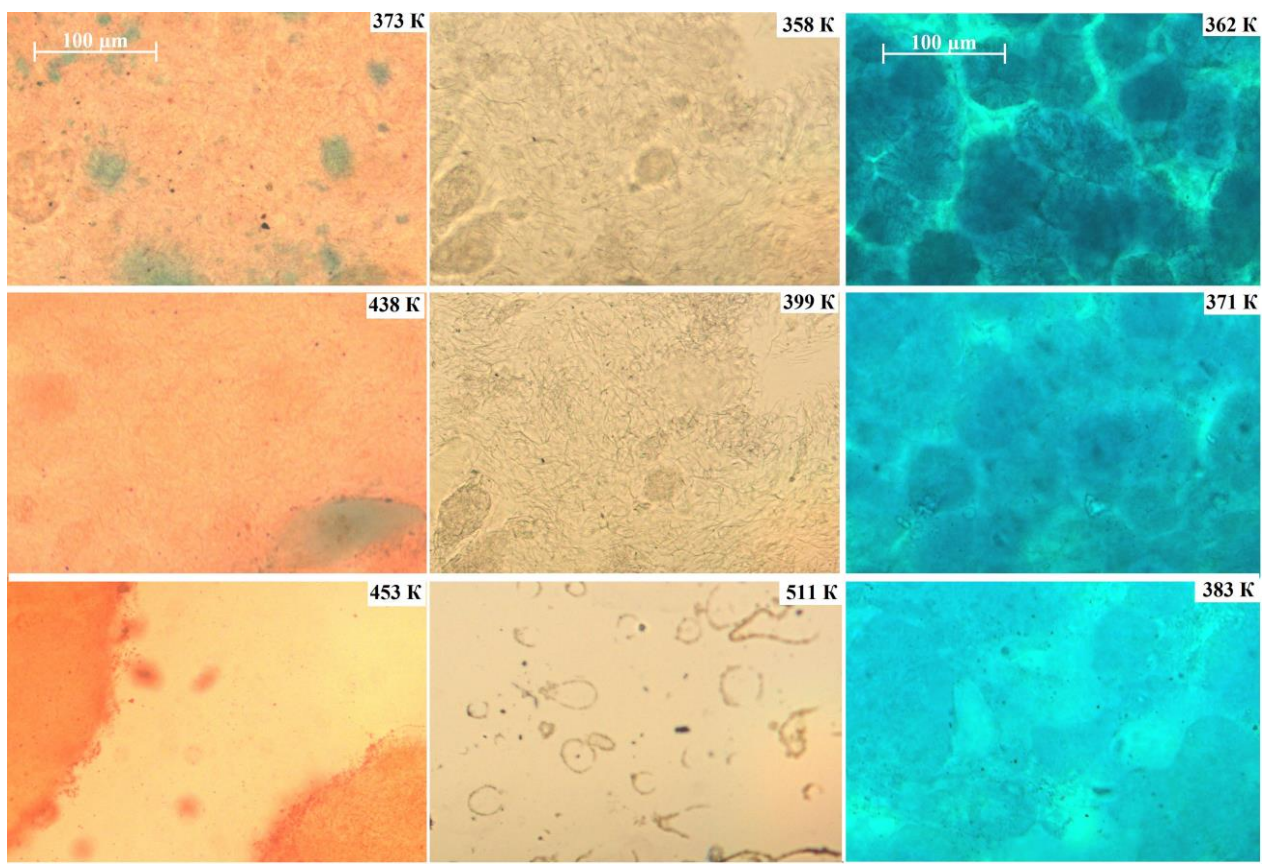

Fig. 4. Microphotographs of the system Litol-24-Cu(II) valerate 5, 10, 20 wt. % (left, middle and right columns respectively) at different temperatures in heating regime.

2.3. Dielectric spectroscopy

To reveal the phase transitions, we also recorded the dielectric spectra (Figs. 5-8) for the standard electric values, resistance, capacity, inductance, with the Fluke PM 6303 device in the frequency range from 100 Hz to 1 MHz at measurement inaccuracy 0.1 % [22]. At frequencies above 1 MHz and at different temperatures, we suggest occurrence of peaks of dissipation factor, which may correspond to the different relaxation processes.

The software of Fluke PM 6303 allows us to control the electric parameters and to record results into ASCII files for further measurement-processing.

The samples in measuring cells with electric capacitance 7 pF are placed into the thermostating block, including the air bath with the attemperator, allowing us to change the heating rate of the samples. Volume of an examined substance in a measuring cell equals 2.12 cm$^3$, the electrode area equals 7.07 cm$^2$. The temperature of the samples inside the thermostat was recorded with the digital thermometer Aktakom (https://www.aktakom.com) ATT-2002 at measurement inaccuracy was not exceeding 0.5º. Measurements of electrical characteristics were carried out in continuous mode and recorded at a given frequency. We provided the heating rate 10º in two successive cycles from room temperature to the melting temperatures of pure Litol-24, cooling was carried out to room temperature, the subsequent heating was carried out to temperatures exceeding the melting temperature by 10-30 K. When measuring the capacitance of the samples, a temperature hysteresis was observed. The reheating data were used to analyze the dielectric spectra.

In dielectric measurements, we applied the ideal capacitor scheme [23], and we recorded the frequency dependencies for the dielectric constants $\varepsilon'$, $\varepsilon''$ and the dissipation factor $tg\delta = 1/\omega CR$.

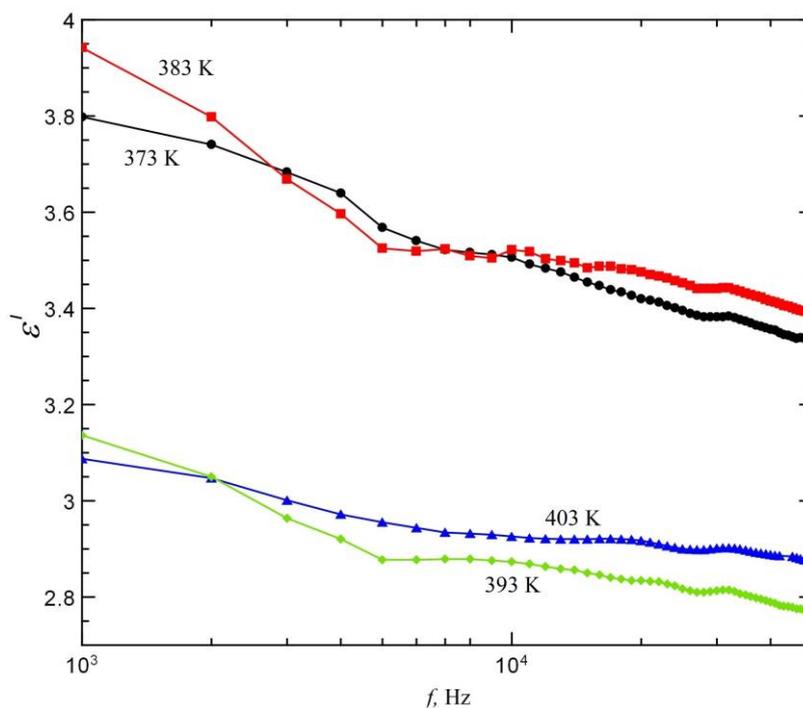

Fig. 5. The frequency dependencies of dielectric permittivity $\varepsilon'$ for Litol-24 with Cu(II) isovalerate at concentration 10 wt. % at temperatures 373-403 K: 373 K (circles), 383 K (square boxes), 393 K (rhombs), 403 K (triangles).

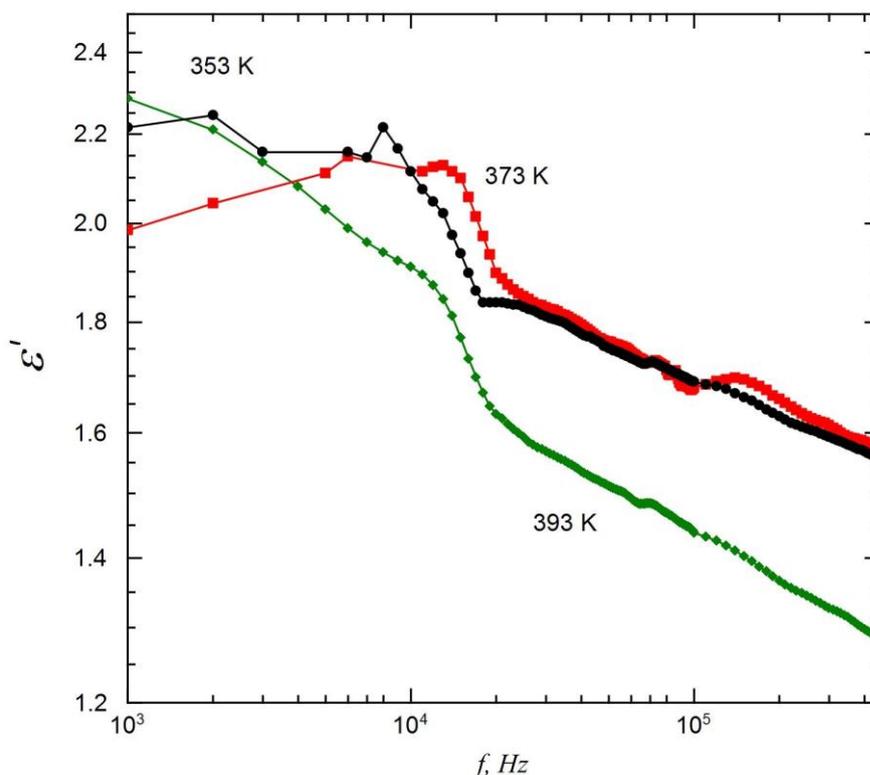

Fig. 6. The frequency dependencies of dielectric permittivity $\varepsilon'$ for Litol-24 with Cu(II) valerate at concentration 10 wt. % at temperatures 353-393 K: 353 K (circles), 373 K (square boxes), 393 K (rhombs).

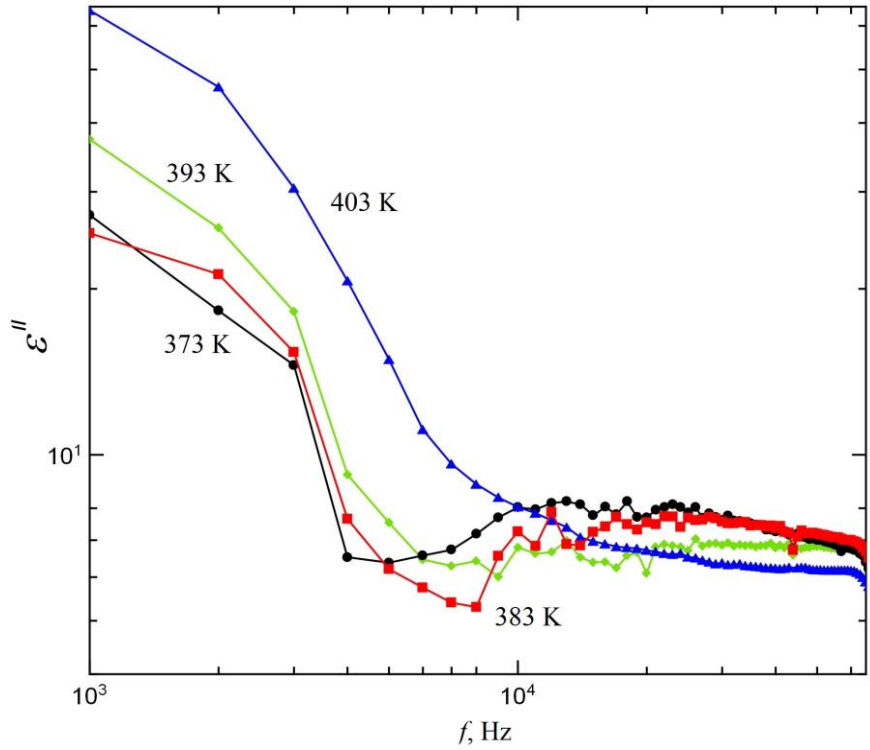

Fig. 7. The frequency dependencies of dielectric permittivity $\varepsilon''$ for Litol-24 with Cu(II) isovalerate at concentration 10 wt. % at temperatures 373-403 K: 373 K (circles), 383 K (square boxes), 393 K (rhombs), 403 K (triangles).

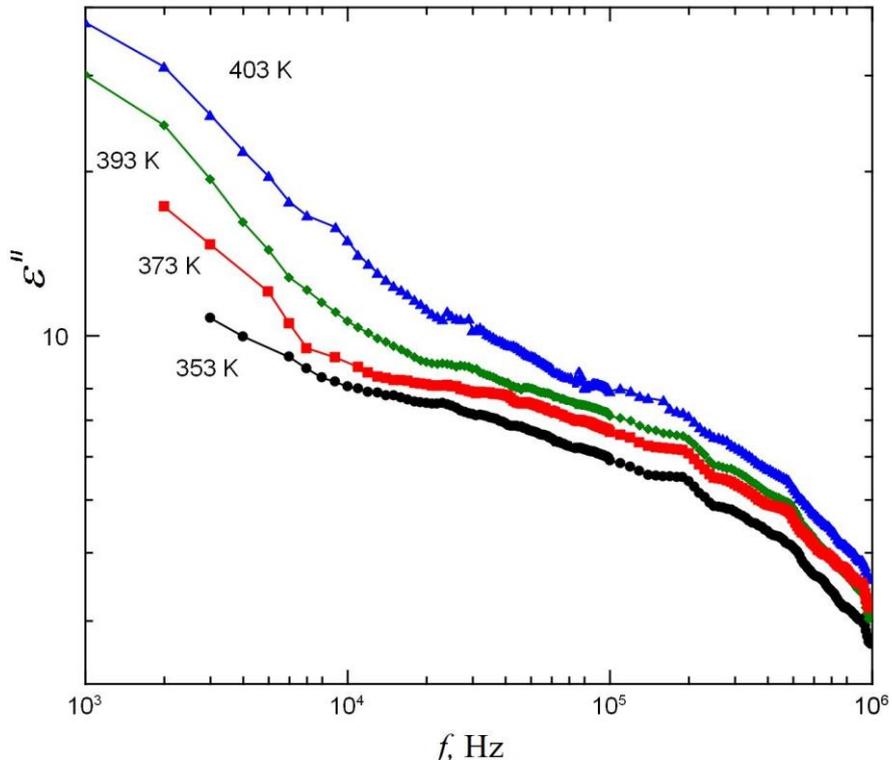

Fig. 8. The frequency dependencies of dielectric permittivity $\varepsilon''$ for Litol-24 with Cu(II) valerate at concentration 10 wt. % at temperatures 353-403 K: 353 K (circles), 373 K (square boxes), 393 K (rhombs), 403 K (triangles).

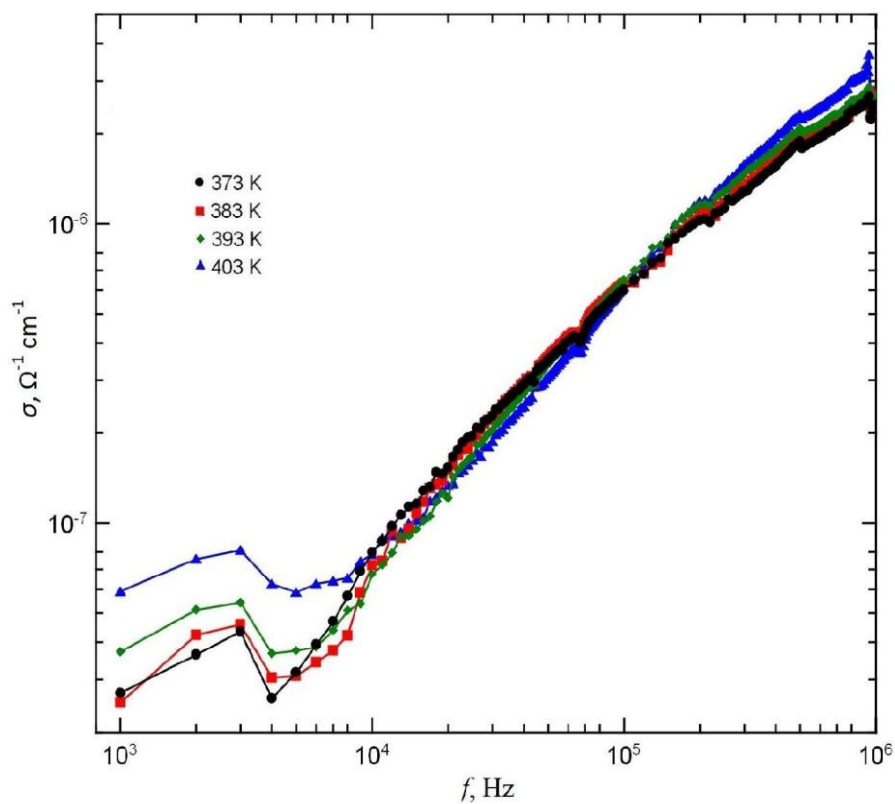

Fig. 9. The frequency dependencies of specific electric conductivity for Litol-24 with Cu(II) isovalerate at concentration 10 wt. % at temperatures 373-403 K: 373 K (circles), 383 K (square boxes), 393 K (rhombs), 403 K (triangles).

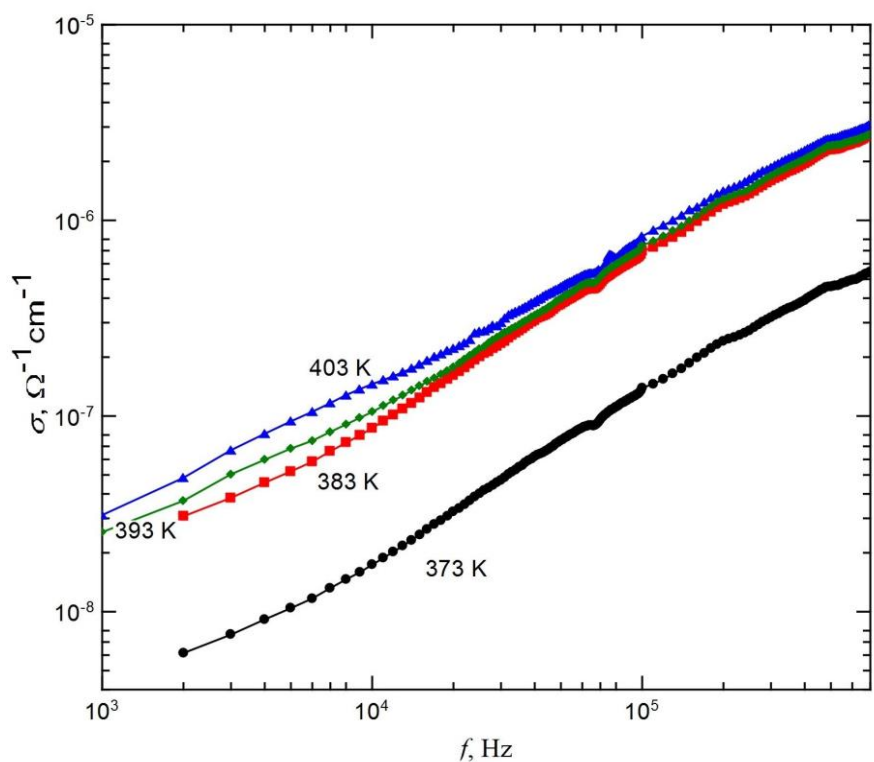

Fig. 10. The frequency dependencies of specific electric conductivity for Litol-24 with Cu(II) valerate at concentration 10 wt. % at temperatures 373-403 K: 373 K (circles), 383 K (square boxes), 393 K (rhombs), 403 K (triangles).

Fig. 9 shows the change in the electric conductivity at 10% isovalerate concentration and at ~7 kHz that may be evidence of the expected *I-Col* phase transition [5]. Though the similar varying character in $\sigma$ has been observed for the system Litol-24-Cu(II) valerates (Fig 10), it may designate on the isotropic to crystalline phase transition.

At Fig. 11, we plotted the dissipation factor for the system Litol-24-isovalerate Cu(II) 10 wt.%, which denotes the phase transitions else. Fig. 12 shows more imperceptible weak peaks of tgδ in Litol-24-valerate Cu(II) composites, they me be caused by own Litol-24 electric conductivity. Fig. 13 confirms the not uniform character of specific electric conductivity at different concentrations of additives and temperatures for the valerate Cu(II)- and isovalerate Cu(II) - Litol-24 systems

So we have to confirm the character of these transformation with the model of effective media [24, 25] basing on image processing in polarization microscopy and the data of dielectric spectroscopy. The effective media approximations allow us to extract the dielectrical characteristics of the volume inclusion fractions in composite materials; applying an approximation of volume spherical isotropic inclusions, we base on the expression

$$\sum_i \delta_i \frac{\sigma_i - \sigma_{eff}}{\sigma_i + (n-1)\sigma_{eff}} = 0, \qquad (1)$$

where $\delta_i$ is the fraction of each component $i$, $\sigma_i$ is their electric conductivity, and $n$ is Euclidean spatial dimension (here $n=3$). For our system, we can estimate $\delta_i$ from polarizing spectroscopy analysis, and to distinguish $\sigma_i$ of pure Litol-24 and of the composite system Litol-24-isovalerate Cu(II) as measured $\sigma_{eff}$.

The results of such calculations shown in Fig. 14 proves the mesogenic nature of the Cu(II) isovalerate additives.

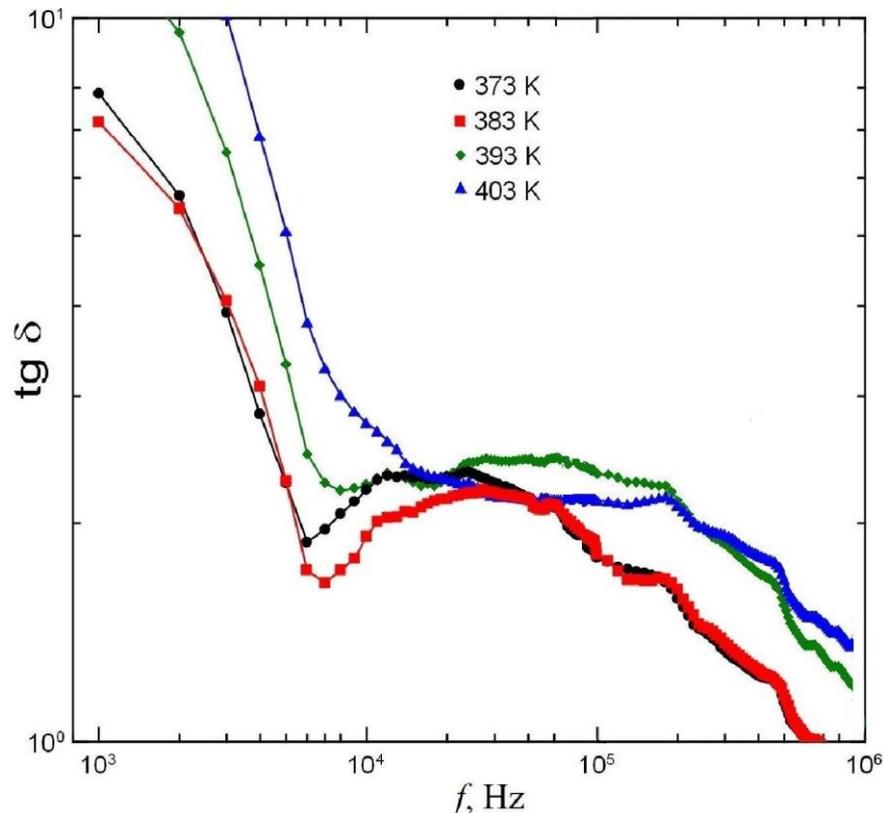

Fig. 11. The frequency dependencies of dissipation factor tgδ for Litol-24 with Cu(II) isovalerate at concentration 10 wt. % at temperatures 373-403 K: 373 K (circles), 383 K (square boxes), 393 K (rhombs), 403 K (triangles).

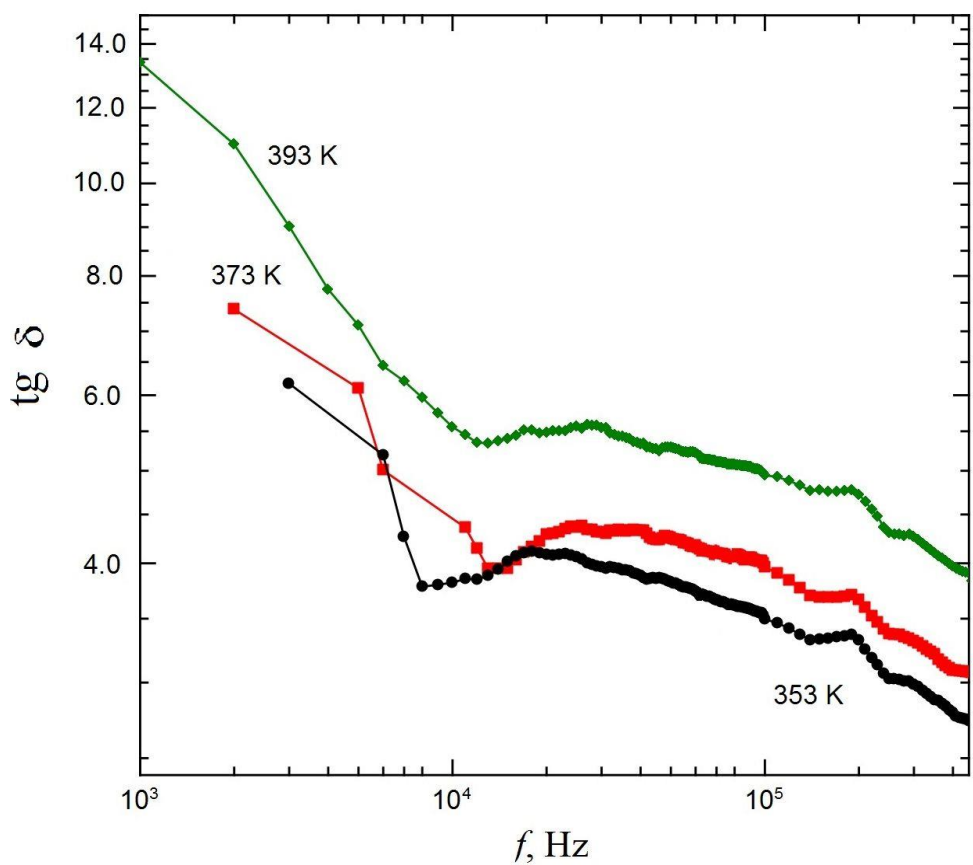

Fig. 12. The frequency dependencies of dissipation factor tgδ for Litol-24 with Cu(II) valerate at concentration 10 wt. % at temperatures 353-393 K: 353 K (circles), 373 K (square boxes), 393 K (rhombs).

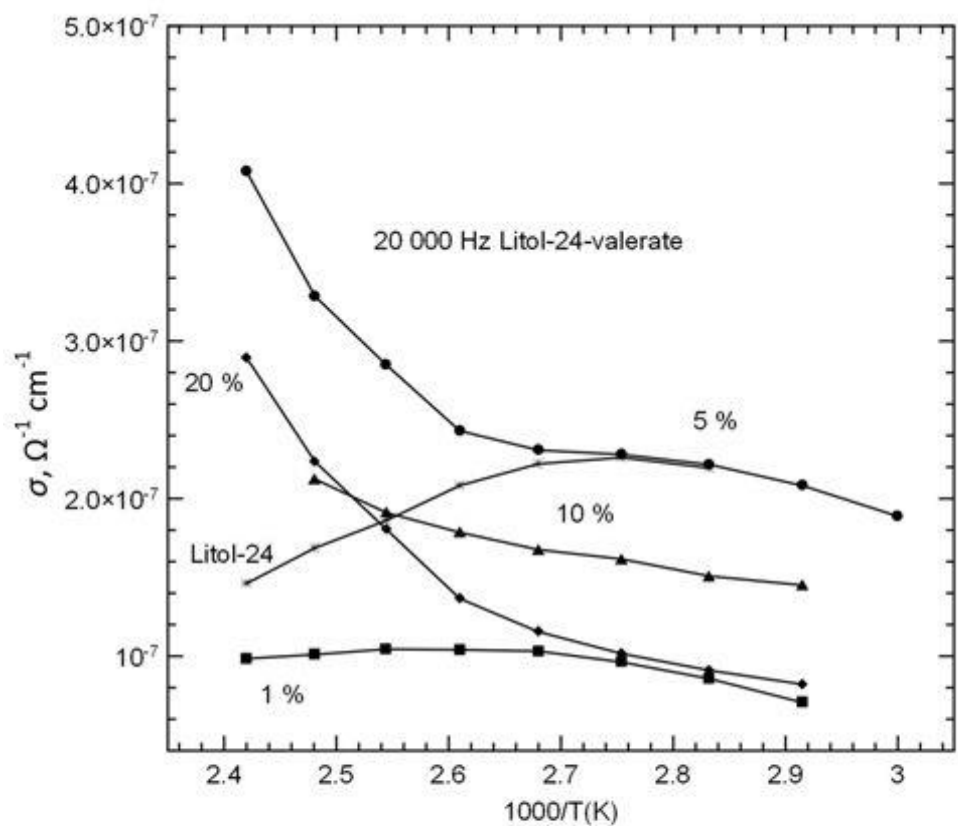

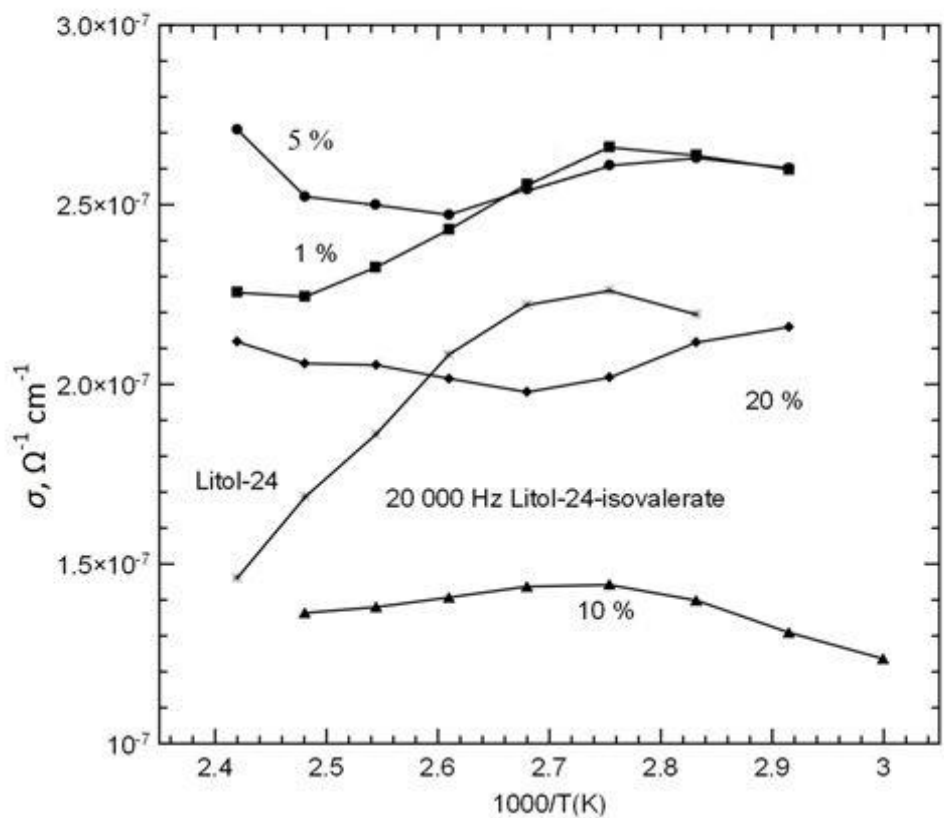

Fig. 13. Specific electric conductivity *vs* temperature for the Litol-24-valerate and Litol-24-isovalerate systems at 20 kHz.

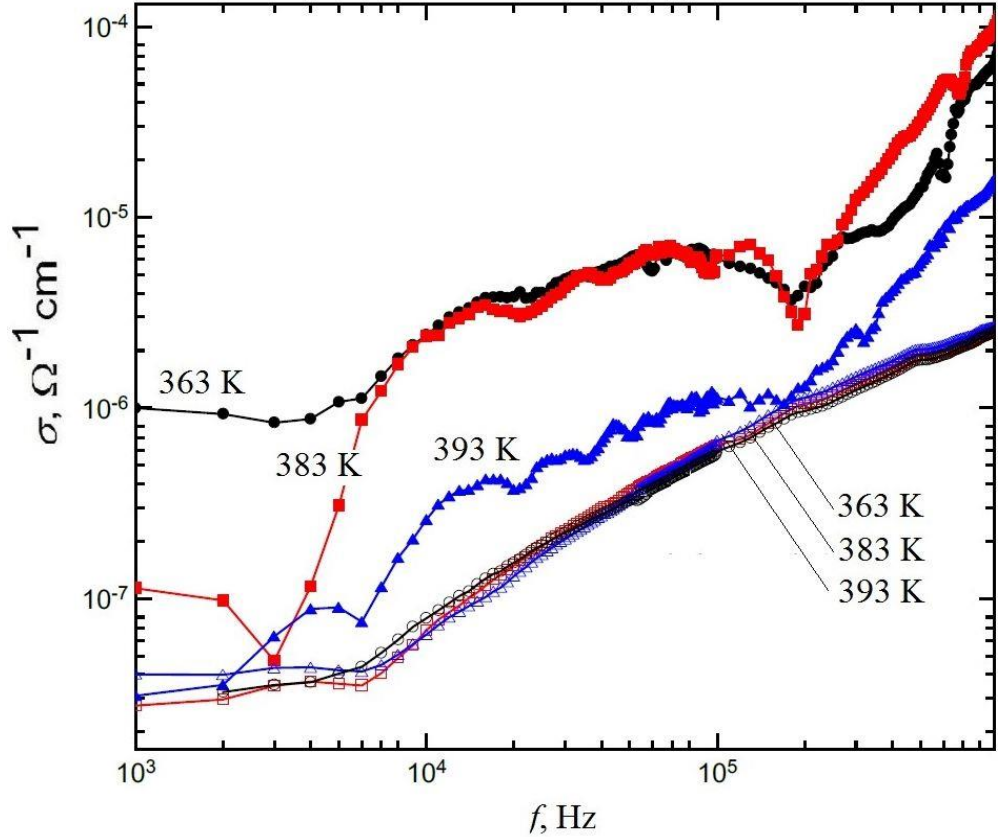

Fig. 14. The frequency dependencies of specific electric conductivity for Cu(II) isovalerate at concentration 10 wt. % at temperatures 363, 383, 393 K: 373 K (black fill circles), 383 K (red square boxes), 393 K (blue triangles) and for the system Litol-24-Cu(II) isovalerate at concentration 10 wt. % (open symbols).

### 3. Modelling

To explain the nature of the phase transitions and to predict the series of possible operating regimes of these lithium greases, we need to compose the theoretical model for the series of the phase transitions between solid and isotropic phases via the columnar phases in melting, which are driven by increase in temperature and insertion of additives into a base lubricant.

As is known [11,12,26], the discotic phases are classified due to the symmetry classes as hexatics (symmetry group $C_6$), 4-fold discotics ($C_4$) and 2-fold discotics ($C_2$). The systems under our consideration form hexatic phases.

Due to the theoretical description [9,10], the systems forming discotic phases have the total phase diagram which is expressed as $Cr \leftrightarrow D \leftrightarrow N_D \leftrightarrow I$.

In this diagram, $Cr$, $D$, $N_D$, $I$ denote a crystalline phase (or the phase of molecular crystal), a discotic phase, a nematic with reverse sign of anisotropy of dielectric and viscoelastic characteristics, and an isotropic phase respectively.

Here we omit the consideration of the transition from the $Cr$ phase $Cr \leftrightarrow D$, as we do not manage with this transformation in tribological conditions.

Physical characteristics of $D$ and $N_D$ phases are symmetrical with respect to the order parameter $\mathbf{n}$: $\psi(-\mathbf{n}) = \psi(-\mathbf{n})$. The $N_D$ phases were not observed in our experiments.

The order parameter of the $D \leftrightarrow N_D$ transition is the multicomponent vector $\mathbf{\psi}$, fixed two-dimension lattice, for the $N_D \leftrightarrow I$ transition, where the order parameter is a uniaxial symmetric second order traceless tensor $Q_{ik}$. The total thermodynamic potential is [11]

$$\Phi = \Phi_1(Q_{ik}) + \Phi_2(\psi) + \Phi_3(Q_{ik}, \psi) + \Phi_4(\chi) + \Phi_5(Q_{ik}, \chi) \quad (2)$$

The potential $\Phi_1(Q_{ik})$ is associated with the $N_D \leftrightarrow I$ transition, here, we will not consider it as unobservable experimentally in the measured temperature range, as well as the $N_D \rightarrow D$ transition, where the Frank moduli $K_{11}$ and $K_{22}$ and correlation length increases.

$$\Phi_2(\psi) = \frac{1}{2}a_2 \sum_{\mathbf{p}} \psi(\mathbf{p})\psi(-\mathbf{p}) + \frac{1}{3}b_2 \sum_{\mathbf{p}_1\mathbf{p}_2\mathbf{p}_3} \psi(\mathbf{p}_1)\psi(\mathbf{p}_2)\psi(\mathbf{p}_3)\delta(\mathbf{p}_1+\mathbf{p}_2+\mathbf{p}_3) +$$
$$\frac{1}{4}c_2 \sum_{\mathbf{p}_1\mathbf{p}_2\mathbf{p}_3\mathbf{p}_4} \psi(\mathbf{p}_1)\psi(\mathbf{p}_2)\psi(\mathbf{p}_3)\psi(\mathbf{p}_4)\delta(\mathbf{p}_1+\mathbf{p}_2+\mathbf{p}_3+\mathbf{p}_4) + D_{\parallel}\sum_{\mathbf{p}} \left| \left(\frac{\partial}{\partial z} + i\delta \mathbf{n}\mathbf{p}\right)\psi(\mathbf{p}) \right|^2$$
$$+ D_{\perp}\sum_{\mathbf{p}} |Q_{\perp}(\psi(\mathbf{p}))|^2 + \Phi_{el}. \quad (3)$$

The term (3) participates in the 2D lattice formation and encloses the gauge terms of the group $Z_2$ in the gradient terms. Here $a_2$, $b_2$, $c_2$, $D_i$, are the coefficients of the Landau expansion, and $K_{ii}$ are the Frank moduli.

The "mixed" term of the thermodynamical potential is

$$\Phi_3 = -\sum_{\mathbf{p}} |\psi(\mathbf{p})|^2 f(Q_{ij}), f(Q_{ij}) > 0, \quad (4)$$

with $f(Q_{ij}) > 0$. If the normalized values of (3) $a_2$, $b_2$, $c_2$ satisfy to the condition $\tilde{f} > \frac{1}{2}\tilde{a}_2 - \frac{1}{9}\frac{\tilde{b}_2^2}{\tilde{c}_2}$

[11], then there is no the nematic $N_D$ between $D$ and $I$ phases.

The other terms $\Phi_4$, $\Phi_5$ of the Landau expansion are responsible for the work of the two-component order parameter $\chi$.

$$\Phi_{el} = \frac{1}{2}K_{11}(\nabla \delta \mathbf{n})^2 + \frac{1}{2}K_{22}(\mathbf{n}_0 \cdot \nabla(\delta \mathbf{n}))^2 + \frac{1}{2}K_{33}(\mathbf{n}_0 \times \nabla(\delta \mathbf{n}))^2. \quad (5)$$

In the notation [7,9], dislocations are described by the elastic term (5), and there are three types of dislocations are possible: the longitudinal edge dislocation (aligned along the columns), the transverse edge dislocation and the screw dislocation (or the Burgers dislocation).

In the BKT theory, the topological dislocation melting of triangular lattice is appeared for finite temperature. So because of discrepancies in electroconductivity curves for composite lithium grease with valerate and isovalerate Cu(II), we assumed as a basis hypothesis, that geometric positions of dislocations correspond to localization of carboxylate additives, and two types of relaxation processes may give evidence of the melting BKT transition $D \rightarrow I$, which we study.

The main result of Kosterlitz and Thouless on the self-consistent equation for a scale dependent electric constant $\varepsilon(r)$ [18] is

$$\varepsilon(r) = 1 + 4\pi^2 y_0^2 K_0 \int_a^r \left(\frac{r'}{a_0}\right)^{4-2\pi U(r')} \frac{dr'}{r'}. \quad (6)$$

Here, for the triangular lattice $K_0^r = K_0^\theta = K_0 = \frac{1}{2\pi}\frac{\mu_0 B_0}{\mu_0 + B_0}\frac{a_0^2}{k_B T}$, $\mu_0$ and $B_0$ are the shear and bulk moduli in absent of dislocations. The triangle lattice is dual to hexagonal one, where the discotic phase is formed.

The values $y_0$ (viz $\ln y_0$) and $a_0$ relate to the core energy and triangular lattice spacing respectively. $U(r')$ is the force of dislocations [28].

In the Kosterlitz-Thouless theory one neglects interactions between dislocations and harmonic lattice vibrations and evaluates the energy of the dislocation system $\mathcal{H}_D$ using continuum elasticity theory. Defining $H_D = -\mathcal{H}_D/k_B T$ we present the Hamiltonian of the dislocations [20]. In the continuum elasticity theory, neglecting interactions between dislocations and harmonic lattice vibrations, one writes [20] the dislocation Hamiltonian in the form

$$H_D = 2\pi \left\{ \sum_{<ij>} K_0^n \mathbf{b}^i \cdot \mathbf{b}^j \ln\left(\frac{r^{ij}}{a_0}\right) - K_0^\theta \left[\frac{(\mathbf{b}^i \cdot \mathbf{r}^{ij})(\mathbf{b}^j \cdot \mathbf{r}^{ij})}{(r^{ij})^2} - \frac{1}{2}\mathbf{b}^i \cdot \mathbf{b}^j\right] \right\} + \ln y_0 \sum_i (b^i)^2, \quad (7)$$

where $H_D$ is normalized to $k_B T$ (the Boltzmann constant and temperature). The vectors $r^{ij}=r^i-r^j$ denote difference in positions of dislocations at the Burgers vector $\Sigma_i b^i = 0$.

Finally, we use the Hamiltonian (7) represented at the lattice variables on prismoidal lattice sites, where an each term describes three different types of dislocations in discotic [20]:

$$H(\{n\}) = 2\pi \sum_{<ij>} b_\mu^i \varepsilon_{\mu\alpha} b_\nu^j \varepsilon_{\nu\beta} M_{\alpha\beta}(\mathbf{r}^{ij}), \quad (8)$$

where $M_{\alpha\beta}(\mathbf{r}^{ij}) = K^r \delta_{\alpha\beta} G(r) + K^\theta (r_\alpha r_\beta / r^2 - \frac{1}{2}\delta_{\alpha\beta})$, $\varepsilon_{\alpha\beta}$ is the antisymmetric Levi-Civita tensor, and $G(r) = \ln(r/a_0)$, here $a_0$ may be considered as a dislocation core radius.

The partition function of (7) is written as

$$Z = \sum_{\{n_\alpha\}} \prod_{n=1}^{6} \left(\frac{1}{n_\alpha!}\right) \int \prod_{i=1}^{N} d^2 r^i \left(\frac{y}{a_0^2}\right)^N e^{H\{(n)\}}. \quad (9)$$

We provided the Monte Carlo simulations with the Metropolis algorithm at the dual to the hexagonal lattice with sizes 48×48×10 and triangular coordinates in the *XY* plane. The temperature parameter $\beta=/k_B T$. The Figs. 15, 16 show results of thermodynamical calculations at $a_0=1$ and at the different "reference" coefficients $K^r=K^\theta=K$ of the order of 1, 2, 3 (Fig. 16, black squares, red circles, and blue triangles respectively) for test of the presence of the phase transitions. At higher values of *K*, the crossover region appears at inverse temperature $\beta > 4$.

We find that the temperature melting 1-order transitions are observed. The increasing in defect terms corresponding of two first summands of (7) leads to the shift in the transition to high temperatures and makes the peaks sharper.

We know that in accordance with the BKT theory [27], dislocations prevent melting of the hexatic phase.

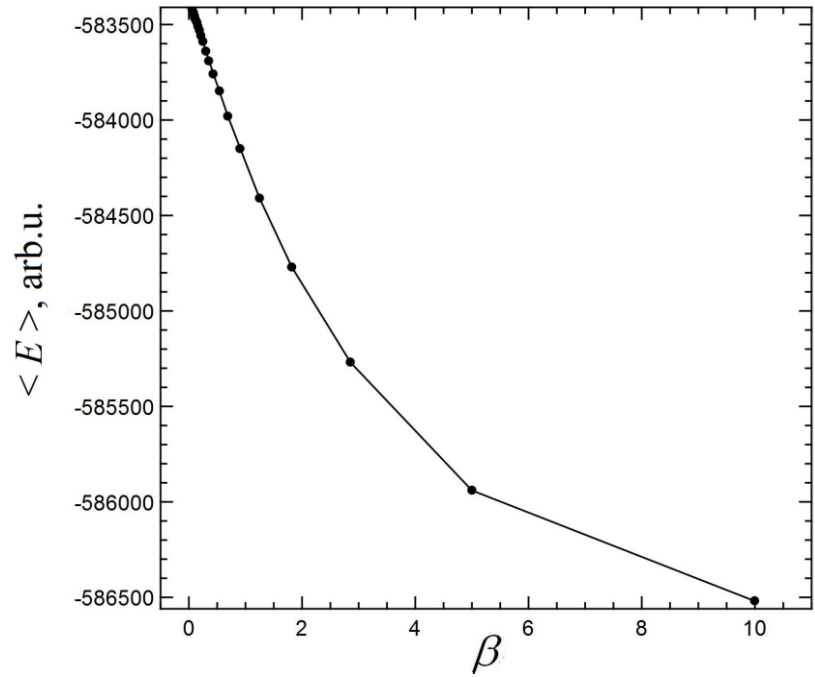

Fig. 15. The temperature dependence of the average energy on dual lattice for the dislocation Hamiltonian (8) plotted with Monte Carlo simulations (errors of calculations ~0.1%); here we use the parameters of the one-constant approximation $K^\tau = K^\theta \sim 1$.

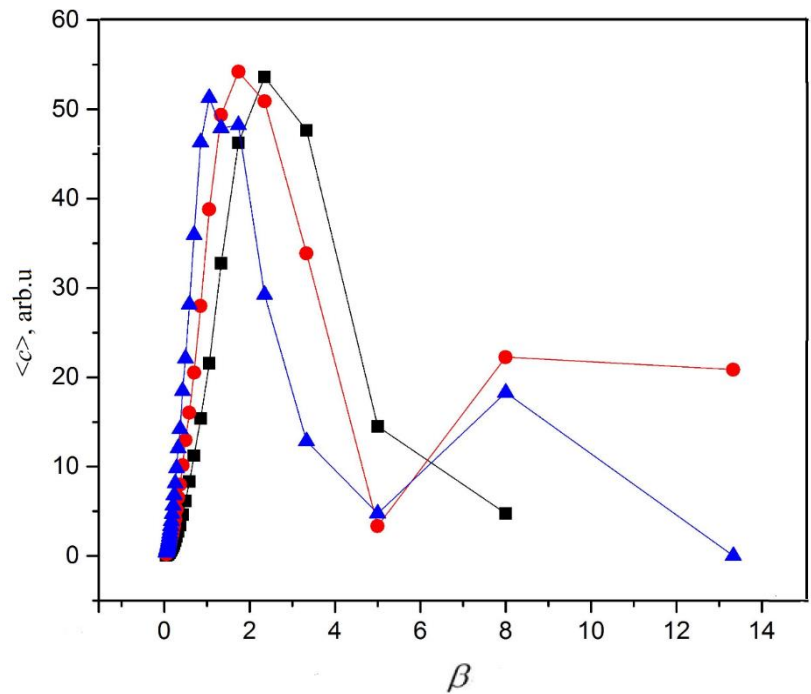

Fig. 16. The temperature dependencies of the specific heat on dual lattice for the dislocation Hamiltonian (7) at different dummy coefficients $K^\tau = K^\theta = K \sim 1$ (squares), 2 (circles), 3 (triangles) plotted with Monte Carlo simulations.

## 4. Discussions and conclusions

We explained the experimentally observed change in character of specific conductivity as belonging to the series of transitions $Cr \leftrightarrow D \leftrightarrow N_D \leftrightarrow I$ in the composite lithium grease doped with the mesogenic additives, which were previously described theoretically [6-14].

In the description of phase transitions in discotics inducing by dislocations, the applied approach BKT is *in medias res* identical to continual theory by Kleman and Oswald [10]. Our lattice version avoids using chemical potentials of the components, so the modelling requires to return the parameters from a dual to hexagonal lattice. In our approach, at the dual lattice, only dislocation thermodynamics may be described, we deal with only model parameters $K$ (in one-constant approximation) and the core radius $a_0$ for dislocation geometry. The energy of the dual lattice Hamiltonian (8) may relate to with the force of dislocations $U(r')$ in formula (6) of the original lattice. However, these theoretical constructions go beyond the scope of this text.

In result of our modelling, we defined the temperatures of 1-st order phase transitions $N_D \leftrightarrow I$, qualitative weak temperature dependencies for shear and bulk moduli ($K$), and the temperature and computational ranges of $K$ before crossover.

Also, the difference between the used BKT model [20] and the Kleman's theory [10] is that they are based on different orders of elasticity theory. However, these approaches operate with the same observables, i.e. elastic constants and the dislocation core radius, so their validity may be experimentally confirmed.

For comparison, deriving the relations between the Frank elastic constants $K$, we can refer to the exotic case of bent core nematics 1Cl-N(1,7)-O6 [29] formed the Col$_t$ phase.

It should be noted, that the phases diagrams with discotic liquid crystals attract a lot attention in connection with their applications in electronics, organic photovoltaics *etc.*, so that different modelling are applicable to predict their operating characteristics.

In certain cases, in the phase diagrams of discotic liquid crystals the lattice Schtokmaer model has been used [30, 31]. In this model, the Schtokmaer potential, so called the "12-6-3" potential, describing a pair interaction of molecules with constant dipole moments, with the exception of the Lenard-Jones potential, includes an additional cubic in inverse distance between molecules term of the dipole interaction. This theory results in qualitative agreement in free energy and allows us to take into consideration molecular conformations of the mesogenic additives in lubricants. But for BKT, this model has distinction in kind.

Finally, we conclude that the BKT model in dual lattice variables may serve to interpret the melting $D \leftrightarrow I$ phase transition and to define it as the first order transition. Depending on composition of the greases, i.e. on the mass fraction of mesogens, corresponded to topological defects, both the thermotropic and lyotropic transitions take place.

## 5. Acknowledgements

The authors are thankful to Prof. O. B. Akopova and Dr. V.V. Terentyev for delivery the lithium greases for dielectric spectroscopy experiments and to Prof. V.V. Belyaev for useful discussions. The measurements on polarization microscopy and dielectric spectroscopy were made using the equipment of the Collaborative Access Center "Center for Polymer Research" of ISPM RAS", and the authors are very grateful to V.P. Chekusova for test operation.